\documentclass[12pt]{article}
\newcommand{\epo}{e^+_1}
\newcommand{\emo}{e^-_1}
\newcommand{\beq}{\begin{equation}}
\newcommand{\eeq}{\end{equation}}
\newcommand{\bea}{\begin{eqnarray}}
\newcommand{\eea}{\end{eqnarray}}

\newcommand{\Emo}{E^-_1}

\newcommand{\omo}{\omega_1}
\newcommand{\pz}{\partial_0}
\newcommand{\po}{\partial_1}
\newcommand{\pzinv}{\frac{1}{\partial_0}}
\newcommand{\pzinvsq}{\frac{1}{\partial^2_0}}

\newcommand{\prop}{\Theta}
\newcommand{\plabel}{\label}
\usepackage{amsmath}
\usepackage{amssymb}

\begin{document}

\begin{titlepage}
\renewcommand{\thefootnote}{\fnsymbol{footnote}}

\hfill TUW-97-10 \\
\begin{center}
\vspace{1cm}

{\Large\bf  Generalized 2d-dilaton models, the true black 
hole and quantum integrability}
\vfill
\renewcommand{\baselinestretch}{1}
\vspace{0.4cm}
{\bf M.O.\ Katanaev$^{1,2,}$\footnotemark[1], 
 W.\ Kummer$^{1,}$\footnotemark[2], 
H.\ Liebl$^{1,}$\footnotemark[3]\\
 and D.V.\ Vassilevich$^{1,3,}$\footnotemark[4]}

\vspace{7ex}

{$^1$Institut f\"ur
    Theoretische Physik \\ Technische Universit\"at Wien \\ Wiedner
    Hauptstr.  8--10, A-1040 Wien \\ Austria}

\vspace{2ex}

{$^2$Steklov Mathematical Institute,\\ Vavilov St. 42, 117966 Moscow, Russia}

\vspace{2ex}
{$^3$Dept. of Theoretical Physics,\\   St. Petersburg University,  
198904 St. Petersburg \\  Russia}
\vspace*{2.5cm}
\vfill

Paper submitted to European Conference on High Energy Physics, 
Jerusalem, August 1997

\footnotetext[1]{e-mail: {\tt katanaev@class.mian.su}}
\footnotetext[2]{e-mail: {\tt wkummer@tph.tuwien.ac.at}}
\footnotetext[3]{e-mail: {\tt liebl@tph16.tuwien.ac.at}}
\footnotetext[4]{e-mail: {\tt vasilev@snoopy.niif.spb.su}}
%  \footnotetext[4]{Permanent address: Department of Theoretical Physics,
%    St. Petersburg University, 198904 St. Petersburg, Russia}

\end{center}
\end{titlepage}

\begin{abstract}

All $1+1$ dimensional dipheomorphism-invariant models can be viewed in
a unified manner.  This includes also general dilaton theories and
especially spherically symmetric gravity (SSG) and Witten's dilatonic
black hole (DBH).  A common feature --- also in the presence of
matter fields of any type --- is the appearance of an absolutely
conserved quantity C which is determined by the influx of matter. 
Only for a subclass of generalized dilaton theories the singularity
structure vanishes together with C.  Such `physical' theories include,
of course, SSG and DBH.  It seems to have been overlooked until
recently that the (classical) 'black hole' singularity of the DBH
deviates from SSG in a physically nontrivial manner.  At the quantum
level for {\sl all} generalized dilaton theories --- in the absence
of matter --- the local quantum effects are shown to disappear.  This
enables us to compute e.g.\ the second loop order correction to the
Polyakov term.  For non-minimal scalar coupling we also believe to
have settled the controversial issue of Hawking radiation to infinity
with a somewhat puzzling result for the case of SSG.
\end{abstract}
\newpage

\section{Introduction}

The prime motivation for investigating generalized dilaton models 
\cite{wit91,ban91,kat86,kat95} and especially 
the  dilaton black hole (DBH), always has been the hope to 
obtain information concerning problems of the 'genuine' 
Schwarzschild black hole (SBH) in d = 4 General Relativity: the 
quantum creation of the SBH and its eventual evanescence because of 
Hawking radiation and the correlated difficulty of information loss 
by the transformation of pure quantum states into mixed ones, black hole 
thermodynamics etc. \cite{alv}. 
On the other hand, essential differences between DBH and SBH  have been
known for a long time. We just quote the Hawking temperature ($T_H$)
and specific heat:
For the DBH $T_H$ only depends on the cosmological constant instead of a 
dependence on the mass parameter as in the SBH.
The specific heat is zero for the DBH and negative for the SBH. 

It is important for any application of the DBH or its generalizations to
compare the respective singularity structure with the ones encountered
in General Relativity (GR), e.g. for the (uncharged) spherically symmetric case.
However, careful 
studies of the singularity structure in such theories seem to be 
scarce. Apart from  \cite{lemos} and our recent work \cite{kat95} we 
are not aware of such a comparison. It always seems to have been 
assumed that the physical features coincide at least qualitatively  
in all respects. During our recent work \cite{kat95} we noted that this is 
not the case: 
{\it For the ordinary dilaton black  hole of
\cite{wit91} 
null extremals are complete at the singularity}. Of course, 
non--null extremals are incomplete, and so at least that property 
holds for the DBH, but, from a physical point of view, it seems a 
strange situation that massive test bodies fall into that 
singularity at a finite proper time whereas it needs an infinite 
value of the affine parameter of the null extremal (describing the 
influx e.g. of massless particles) to arrive. 
This obviously contradicts  Penrose's 1965 theorem \cite{penrose}
which is valid in d=4. Problems related to the application of that 
theorem in d=2 have been voiced also some time ago \cite{christensen} 
within a particular 2d model. 
Thus, from the point 
of view of the SBH care has to be taken within any effort
to extract theoretical insight from the usual DBH, whenever the singularity
itself is involved. This is certainly the case for the last stages of an 
evaporating black hole. 
Thus it is not obvious that 
other physical questions such as e.g. the information paradox
can be dealt with satisfactorily within the DBH model of \cite{wit91}.
Of course, effects related to the horizon alone are not affected 
by our analysis, as long as the horizon is sufficiently far from the related 
singularity. 
\newline

In order to pave the way for a more realistic modelling of the SBH 
we (Section 2) consider a two parameter family of generalized 
dilaton theories which interpolates between the DBH and other models, 
several of whom have been suggested already in the literature 
\cite{lemos,lau,mignemi,fabri}. 
The Eddington--Finkelstein (EF) form of the line element, appearing 
naturally in 2d models when they are expressed as 'Poisson--Sigma 
models' (PSM) \cite{very} is very helpful in this context. 
We indeed find large ranges of parameters for which possibly more
 satisfactory BH models in $d=2$ may be obtained. 

Section 3 is devoted to complete quantum integrability, whereas 
Hawking radiation is treated in Section 4. 

In order to be able to compare the family of dilaton theories 
considered below, the EF metric
\begin{equation}
\plabel{ds2}
(ds)^2= d \bar{v}(2d\bar{u}+l(\bar{u})d\bar{v})
\end{equation}
is most useful, which  explicitly depends on the norm $l = k^\alpha 
k_\alpha$ of the Killing vector $\partial / \partial\bar v$.

Eq.\ (\ref{ds2}) 
is particularly convenient to make contact with the PSM formulation 
which can be obtained for all covariant 2d theories \cite{alt}. They 
may be summarized in a first order Palatini type action

\begin{equation}
\plabel{lagr}
L=\int X^+T^- +X^-T^+ +Xd\omega -e^- \wedge e^+V(X)
\end{equation}

In our present case only vanishing torsion

\begin{equation}
\plabel{tors}
T^{\pm}=(d \pm \omega)e^{\pm}
\end{equation}
as implied by Eq. (\ref{lagr}) is expressed in terms of light--cone (LC) 
components for the zweibein one form $e^a$ and for the spin 
connection one form ${\omega^a}_b = {\epsilon^a}_b \omega$ . The 
`potential' V determines the dynamics. It is simply related to the 
Killing norm $l$ in the EF gauge because (\ref{lagr}) can be solved exactly
for any integrable $V$ \cite{alt} with the solutions (constant curvature is
excluded)

\begin{eqnarray}
\plabel{zweib}
e^+=X^+df \\ e^-=\frac{dX}{X^+}+X^-df
\end{eqnarray}

A similar equation for $\omega$ will not be needed in the following.
The line element immediately yields the EF form (\ref{ds2}) with $\bar{u}=X$
and $\bar{v}=f$. The Killing norm

\begin{equation}
\plabel{kill}
l=C-\int^X V(y)dy
\end{equation}
follows from a conservation law

\begin{equation}
\plabel{const}
C=X^+X^- +\int^X V(y)dy
\end{equation}
common to all 2d covariant theories \cite{very} \cite{alt}
which is related to a 
global nonlinear symmetry \cite{kumwid}. 
The usual dilaton models are produced by the 
introduction of the dilaton field $\phi$ in $X = 2 \exp 
(-2\phi)$, together with a conformal transformation $e^a=exp(-\phi)\tilde{e}^a$ 
of $d\omega$ in (\ref{lagr})

\begin{equation}
\plabel{eps}
\epsilon^{\mu \nu} \partial_{\mu} \omega_{\nu} =-\frac{R \sqrt{-g}}{2}
\end{equation}
with the components $\omega_\nu$ expressed by the vanishing of the 
torsion (\ref{tors}) in terms of the zweibein.

The relation
\begin{equation}
\label{change}
\sqrt{-g}R=\sqrt{-\tilde{g}}\tilde{R} + 2\partial_{\mu} (\sqrt{-\tilde{g}}
{\tilde g}^{\mu\nu} \partial_{\nu} \phi) 
\end{equation}
will be used frequently. 

Let us consider the SBH in a little more detail. The starting point is
the Schwarzschild solution in EF coordinates \cite{wald}

\begin{equation}
\label{ef}
ds^2=2dvdr + \left( 1-\frac{2M}{r} \right)dv^2 -r^2d {\Omega}^2,
\end{equation}
whose $r-v$ part is of the type (\ref{ds2}).
Thus the radial variable
may be identified with $\bar{u}$ in (\ref{ds2}). 
Indeed the correct singularity behavior is obtained from the 
PSM action (\ref{lagr}) with

\begin{equation}
\label{v2}
V=-\frac{M}{X^2}.
\end{equation}

The 'pure' PSM model  for the SBH with potential (\ref{v2})
is fraught with an important drawback: When 
matter is added the conserved quantity $C$ 
in (\ref{simcon}) simply generalizes 
to a similar conserved one with additional matter contributions. 
As shown in the second reference of \cite{kumwid}, from the equations of motion
the conservation law then refers to 
\begin{equation}
C \to C + C^{(m)}
\end{equation}
where $C^{(m)}$ vanishes in the absence of (bosonic as well as fermionic)
matter. It should be emphasized that $C$ coincides with the mass parameter
in the ADM as well as Bondi sense \cite{admbondi}
for the DBH and for spherically symmetric gravity up to numerical factors,
as has been analyzed in detail in \cite{lieblvassil}.
In \cite{kumwid} it was also pointed out that the definition of such 
a conserved quantity does not require an asymptotically flat space-time.
Thus even before a BH is formed by the 
influx of matter an 'eternal' singularity as given e.g. by (\ref{v2}) for 
the SBH, is present  in which the mass $M$ basically cannot 
be modified by the additional matter. 
A general method to produce  at the same time 
a singularity--free ground 
state with, say, $C = 0$ is provided by a  Weyl 
transformation of the original metric. It simply generalizes what 
is really behind the well-known construction of the DBH theory. 
Consider the transformation 
\begin{equation}
\label{91}
\tilde{g}_{\mu \nu} =\frac{g_{\mu \nu}}{w(X)}
\end{equation}
in (\ref{ds2}) with (\ref{kill}) together with a transformation of $X$

\begin{equation}
\label{92}
\frac{d X}{d \tilde{X}} =w(X(\tilde{X})).
\end{equation}
This reproduces the metric $\tilde{g}_{\mu \nu}$ in EF form
\begin{equation}
\label{93}
(ds)^2=2df \left( d\tilde{X} +\left(\frac{C}{w}-1 \right)df \right)
\end{equation}
with a flat ground--state $C = 0$.  Integrating out
$X^+$ and $ X^-$ 
in (\ref{lagr}), and using the identity  (\ref{change})
with $\phi = \frac{1}{2} \ln w$
one arrives at a generalized dilaton theory

\begin{equation}
\label{95}
\it{L}=\sqrt{-\tilde{g}} \left( \frac{X}{2}R+\frac{Vw}{2}\tilde{g}^{\mu \nu}
\partial_{\mu} \tilde{X} \partial_{\nu} \tilde{X} -Vw \right)
\end{equation}
where $X$ is to be re-expressed by $\tilde X$ through the integral 
of (\ref{92}).  

It should be noted that the (minimal) coupling to matter is 
invariant under this redefinition of fields. Clearly (\ref{95}) is the most 
general action in $d=2$  where the flat ground state
corresponds to $C=0$. $V(X)$ may determine an 
arbitrarily complicated singularity structure. 
The DBH is the special case $V=\lambda^2=const.$ Then $\tilde{X}$ is easily seen
to be proportional to the dilaton field. The SBH results from the choice 
$V=X^{-1/2}.$ Using (\ref{92}) and comparing (\ref{93}) with (\ref{ef})
in that case with the interaction constant in $w$ fixed by 
$w=\frac{\tilde{X}}{2}$,
the conserved quantity $C$ is identified with the mass $M$ of the BH and 
(\ref{95}) turns into the action of spherically reduced 4D general relativity
\cite{lau}. Unfortunately such a theory cannot be solved exactly if coupling
to matter is introduced.

\section{Dilaton Models with Schwarzschild-like
 Black Holes}

In \cite{kat97} all models with one horizon and power type singularity 
were analyzed globally. They are described by the action

\begin{equation}
\plabel{ldil}
L=\int d^2x \sqrt{-g}e^{-2\phi}(R+4a(\nabla\phi)^2 +Be^{2(1-a-b)\phi})
\end{equation}

This action covers e.g. the CGHS model \cite{wit91}
for $a=1$, $b=0$, spherically 
reduced gravity \cite{lau} $a=\frac{1}{2}$, $b=-\frac{1}{2}$, 
the Jackiw-Teitelboim model \cite{jackiw} $a=0$, $b=1$.
Lemos and Sa \cite{lemos} give the global 
solutions for $b=1-a$ and all values of $a$  
Mignemi \cite{mignemi} considers
$a=1$ and all values of $b$. The models of \cite{fabri} correspond to $b = 
0, a \leq 1$. It turns out \cite{kat97} that the region leading to 
Penrose diagrams like the one of the genuine black hole is 
restricted to the range: 
\begin{eqnarray}
I) & b < 0; & a < 1 \\
II) & b < 1-a, & 1 \leq a < 2, 
\end{eqnarray}

where region II) is fraught with null-completeness at the 
singularity. The straight line $ b = a -1$ in I) describes the 
'physical' theories with vanishing singularity at ${\cal C} = 0$. 
Although $b = 0$ (except for the DBH at $a = 1$!) has vanishing 
curvature asymptotically, these theories are in that range of 
Rindler type unfortunately. 

\section{Quantum theory of 2d-models}

\subsection{Quantum integrability of matterless theories}

Stimulated by the `dilaton black 
hole' \cite{wit91,ban91} numerous studies of quantized gravity 
in the simplified
setting of 2D models were performed 
\cite{louis}-\cite{brs}. 
Louis-Martinez et al.\ \cite{louis} treated generic 2D dilaton gravity
in the second order formalism
\begin{equation}
 \plabel{lbegin}
{\cal{L}}_{(1)}=\sqrt{-g}
\left(-X\frac{R}{2}-\frac{U(X)}{2}(\nabla X)^2+V(X))\right)
\end{equation}
using a Dirac quantization scheme.
A gauge theoretical formulation for string inspired gravity was developed and
quantized by Cangemi and Jackiw \cite{cangemi-jackiw}. In ref.\ 
\cite{benedict} 
their solutions were shown to be equivalent to the ones of \cite{louis}.
A Dirac approach was recently used to quantize string inspired dilatonic 
gravity \cite{KuRoVa}. In an alternative
approach spherically symmetric gravity was quantized in Ashtekar's
framework by Kastrup \cite{kastrup} and in a geometrodynamical
formulation by Kucha\v{r} \cite{kuchar}.  In particular Strobl
\cite{stroblquant} has treated a large class of 2D gravity theories
within the Poisson-Sigma approach.  
A common feature of all these studies is that due to the particular structure
of the theory the constraints can be solved exactly, yielding a finite
dimensional phase space. Then as a consequence of Dirac quantization
it is found that quantum effects for only a finite number of 
variables are observed. Physically this is in agreement with the fact 
that dilatonic gravity describes no propagating gravitons.
Due to the particular structure of the theory the constraints can be solved
exactly yielding a finite dimensional reduced phase space.
This remarkable property raises hope that in the case of
dilatonic gravity one will be able to get insight into the information
paradox without being forced to deal with the ultraviolet problems of higher dimensional
gravities. Of course, despite of its many appealing features this approach by
itself is insufficient to describe Hawking radiation in quantum
gravity. In the presence of an additional matter field again an infinite
number of modes must  be quantized. In order to tackle that 
problem the results mentioned above first 
of all should be translated into the language of (non 
perturbative) quantum field theory described by the path integral 
as the most adequate method for dealing with infinite dimensional 
quantum systems.

Indeed, one--loop quantum
corrections to the classical action and renormalization group equations 
have been also considered perturbatively \cite{odintsov}.
Matter fields are easily included in this approach. 
In this way, however, even 
pure dilatonic gravity (\ref{lbegin}) was found to exhibit a highly 
non--trivial
renormalization structure, undermining the main motivation for
considering dilatonic gravity as a simple toy model of quantum
black hole physics! 
But even more serious, in our opinion, is the contradiction of these
results with the ones from Hamiltonian approaches as mentioned in the last
paragraph.

Thus a formalism is required which would combine integrability
and simple ultra violet properties of the reduced phase space
quantization with the possibility to include matter and obtain 
the local quantities of the field theoretical quantization.

We remove this contradiction by demonstrating that 
 in pure
dilatonic gravity (\ref{lbegin}) there are no local quantum
corrections in the effective action for the path integral 
approach as well.
To this end we generalize \cite{kumschwarz,kumhaid} and
perform an exact non-per\-tur\-ba\-tive path integral
quantization of a generic 2D dilaton model containing {\it all} the above models.
We give the explicit form of the generating functional for connected Green
functions. Adding matter fields in general destroys  the 
functional integrability
and suffers therefore from the same weaknesses as the first approach.
However, the particular case of JT gravity \cite{jackiw}  even
in the presence of matter fields allows an exact path integral
quantization\footnote{For the matterless case this model was quantized
exactly already by Henneaux \cite{henneaux}.}.

Main technical features of our approach are the use of the first order
action for Cartan variables in the temporal gauge, corresponding 
to an Eddington Finkelstein (EF) gauge for the 
metric \cite{kumschwarz}. 
Our analysis is local, meaning that
we assume asymptotic fall off  conditions for all fields. This is
enough for the first step of a path integral quantization, which 
of course, in a second step should be adapted to take 
into account global effects familiar from  the reduced phase space approach.

We first show the quantum equivalence of the second order form
(\ref{lbegin}) to the first order action 
\begin{equation}
  \plabel{lfirst}
  {\cal{L}}_{(2)}=X^+De^-+ X^-De^++Xd\omega +\epsilon(V(X)+X^+X^-U(X)) ,
\end{equation}
where $De^a=de^a+(\omega \wedge e)^a$ is the torsion two form,
the scalar curvature
$R$ is related to the spin connection $\omega$ by $-\frac{R}{2}=* d\omega$
and $\epsilon$ denotes the volume two form
$\epsilon=\frac{1}{2}\varepsilon_{ab}e^a \wedge e^b=d^2x   \det
e_a^{\mu}=d^2x\,(e)$.  Our conventions are determined by $\eta=diag(1,-1)$
and $\varepsilon ^{ab}$ by $\varepsilon ^{01}=-\varepsilon ^{10}=1$. We
also have to stress that even with Greek indices $\varepsilon^{\mu \nu}$
 is always understood as the antisymmetric symbol and never as the
corresponding tensor. 
The generating functional for the Green functions is given by
\begin{equation}
\plabel{wfirst}
 W=\int ({\cal D}X)({\cal D}X^+)({\cal D}X^-)({\cal D}e^a_{\mu})_{gf}
({\cal D}\omega_{\mu})\exp
\left[i\int_x {\cal{L}}_{(2)} +{\cal{L}}_s \right]\; ,
\end{equation}
where ${\cal{L}}_s$ denotes the Lagrangian containing source terms for the
fields. However, since dilaton gravity does not  any dependence on
$X^{\pm}$ and on $\omega_{\mu}$ we do not introduce the corresponding sources at
this point.
A suitable gauge fixing
is $e_0^-=e_1^+=1$, $e_0^+=0$. It is easy to check that in the following
no division by $e_0^+$ needs to be performed. Also note that
$\det  g = \det  e =1$.
Performing the functional integration with respect to  $\omega_0$ and $\omega_1$ results in
\begin{equation}
\plabel{w2}
 W=\int ({\cal D}X)({\cal D}X^+)({\cal D}X^-)({\cal D}e^a_{\mu})
\delta_{\omega_0}\delta_{\omega_1}\exp
\left[i\int_x \hat{\cal{L}}_{(2)} +{\cal{L}}_s \right] .
\end{equation}
The path integral measure in our gauge is
\begin{equation}
({\cal D}e^a_{\mu})_{gf}=F_{FP}{\cal D}e_1^- \label{degf}
\end{equation}
where $F_{FP}$ is the Faddeev--Popov factor.
We use the abbreviations
\begin{eqnarray}
  \delta_{\omega_0}&=&\delta \left(\partial_1X-X^+e_1^-+X^-e_1^+ 
  \right)\; ,  \\
 \delta_{\omega_1}&=&\delta \left(-\partial_0X+X^+e_0^--X^-e_0^+ 
 \right)\; ,   \\
\hat{\cal{L}}_{(2)}&=&\varepsilon ^{\mu\nu}
\left[X^+\partial_{\mu}e_{\nu}^-+X^-\partial_{\mu}e_{\nu}^++e_{\mu}^+e_{\nu}^-
\left(V(X)+X^+X^-U(X)\right)\right]\; .
  \plabel{simcon}
\end{eqnarray}
Integration over $X^+$ and $X^-$ finally yields
\begin{equation}
  \plabel{wequivalent}
  W=\int ({\cal D}X)({\cal D}g_{\mu \nu})_{gf} \exp
\left[i\int_x {\cal{L}}_{(1)} +{\cal{L}}_s \right]
\end{equation}
In terms of $g_{\mu\nu}$ our gauge condition becomes the
Eddington--Finkelstein gauge $g_{00}=0$, $g_{01}=1$ with the single
unconstrained component $g_{11}=e_1^-$. The path integral measure
becomes
\begin{equation}
({\cal D}g_{\mu \nu})_{gf}=F_{FP}{\cal D}g_{11} \label{dggf}
\end{equation}
with the same Faddeev--Popov determinant as before (\ref{degf}). As a result
of the commonly used introduction of that determinant in our gauge $F_{FP}$
even turns out to be field independent.
${\cal{L}}_{(1)}$ is exactly given by (\ref{lbegin}). To obtain
(\ref{wequivalent}) we used $(e) \equiv \sqrt{-g}$, $(e)^2g^{\alpha\beta}=
\varepsilon ^{\alpha\gamma}\varepsilon ^{\delta\beta}g_{\gamma\delta}$
and the relation
\begin{equation}
  \plabel{omega}
\tilde{\omega}_{\mu}=\eta_{ab}\frac{\varepsilon^{\alpha\beta}}{(e)}
e_{\mu}^a\partial_{\alpha}
e_{\beta}^b \qquad ,
\end{equation}
the tilde indicating the special case of vanishing torsion
such that in (\ref{lbegin}) 
\begin{equation}
\plabel{Rchap6}
\sqrt{-g}\frac{R}{2}=\varepsilon^{\nu \mu}\varepsilon^{\alpha \beta}
\partial_{\mu}\left(\frac{e_{\nu}^a}{e} \partial_{\alpha} e_{\beta a}\right).
\end{equation} 
Therefore, (\ref{omega}) will only produce the torsionless part of the scalar 
curvature as 
it is given in conventional dilaton theories.
Of course, an additional conformal transformation of the zweibein
would result in additional kinetic terms in the Lagrangian.

Thus the quantum theory of (\ref{lfirst}) is indeed equivalent to the one from
the action (\ref{lbegin}).

In our quantization program we use \cite{kumhaid} 
the canonical BVF \cite{brs}
approach in order to obtain the determinants that appear by fixing the gauge
in (\ref{lfirst}).
We will be working in a `temporal' gauge
which corresponds to an Eddington Finkelstein gauge for the metric defined by:
\begin{equation}
  \plabel{gaugefix}
  e_0^+ = \omega_0=0 \quad , \quad e_0^-=1
\end{equation}

After computing the extended Hamiltonian by the introduction of 
the usual two types of ghosts for a stage 1 Hamiltonian, 
following the steps of \cite{kumhaid} we finally arrive at the
generating functional for the Green functions is

\begin{equation}
  \plabel{gen}
  W=\int ({\cal D}X)({\cal D}X^+)({\cal D}X^-)({\cal D}e^+_1)({\cal D}e^-_1)({\cal D}\omega_1)F\exp
\left[i\int_x {\cal{L}}_{(2)} +{\cal{L}}_s \right] ,
\end{equation}
where $F$ denotes the determinant 

\begin{equation}
F=\det (\delta_i^k \partial_0 +{\cal C}^k_{i2} )=
(\det \partial_0 )^2 \det (\partial_0 + X^+U(X)).
\plabel{FPdet}
\end{equation}

${\cal L}_{(2)}$ is the gauge fixed part of the Lagrangian (\ref{lfirst}) and
${\cal L}_s$ denotes the contribution of the sources:
\begin{equation}
  \plabel{source}
  {\cal{L}}_s = j^+e_1^- + j^-e_1^+ + j\omega_1 +J^+X^- +J^-X^+ +JX
\end{equation}

Contrary to the standard approach to the path integral we now 
integrate {\sl first} the 'coordinates' $e_1^\pm, \omega_1$ and 
use the resulting $\delta$-functions to perform the 
$X$-integrations yielding the generating functional 
of connected Green functions
\begin{equation}
  \plabel{Z}
  Z=-i \ln W =\int JX +J^-\frac{1}{\partial_0}j^+ +
J^+ \frac{1}{\partial_0 +U(X)\frac{1}{\partial_0}j^+}
\left(j^- -V(X)\right),
\end{equation}
where $X$ has to be replaced by
\begin{equation}
  \plabel{X}
  X=\frac{1}{\partial_0^2}j^+ +\frac{1}{\partial_0}j \qquad .
\end{equation}
It should be stressed that the determinant $F$ is precisely canceled by the
determinants appearing during these last three integrations.
Eq.\ (\ref{Z}) gives the exact non-perturbative generating functional
for connected Green functions and it does not contain any 
divergences, because it clearly describes tree--graphs only. 
Hence no quantum effects remain. 

Now we turn to the ill defined expressions $(\partial_0)^{-1}$.
As shown in \cite{KLV3} a proper (vanishing) asymptotic behavior results
from a regularization 
($\mu=\tilde{\mu}-i\varepsilon, 
\lim_{\mu \to 0}:=\lim_{\tilde{\mu} \to 0}\lim_{\varepsilon \to 0}$)
\begin{equation}
\plabel{ifcut}
\pz^{-1} \Rightarrow
\begin{cases}
 \lim_{\mu \to 0}\left(\pz -i\mu \right)^{-1}=
 \lim_{\mu \to 0}\left(\nabla_0^{-1} \right) \\
 \lim_{\mu \to 0}\left(\pz +i\mu \right)^{-1}=
 \lim_{\mu \to 0}\left(\tilde{\nabla}_0^{-1} \right) \quad
\end{cases}
\end{equation}
where $\mu^2$ insures a proper infrared cutoff.
Since each partial integration above involved either $X$, $X^+$ or $X^-$
which in turn all exhibit at least a $\partial_0^{-1}$ behavior,
all our previous steps are justified.

The theory thus produces tree graphs only. It can also be shown 
that (in our gauge!) the effective action reduces to the 
classical one --- for the local quantization used here. As shown 
in \cite{kum96} the Jackiw--Teitelboim model may be even quantized 
exactly in this case even when matter is present, although this 
methods fail (in the present version) for more complicated 2d 
theories.

\subsection{Two-loop matter effects}

What can be done in the presence of matter 
is to consider  perturbation theory in the 
{\it matter field},
treating the geometrical part still exactly by a {\it nonperturbative} 
path integral \cite{KLV3}. Our approach thus differs 
fundamentally from the conventional
`semiclassical' one \cite{wit91} in which (mostly only one loop) effects of
matter are added and the resulting effective action subsequently is solved
classicaly.

We take, as our starting point, the action for 1+1
gravity to be the spacetime integral over the Lagrangian 
\begin{equation}
{\cal{L}}={\cal{L}}^g + {\cal{L}}^m +{\cal{L}}^s \quad ,
\end{equation}
which is a sum of the gravitational, the matter and a source contribution.

Our matter contribution is a minimally coupled scalar field whose 
Lagrangian ${\cal{L}}^m$ is given by 

\begin{equation}
\plabel{lmatter}
{\cal{L}}^m = \frac12 \sqrt{-g}g^{\mu \nu}\partial_{\mu}S \partial_{\nu}S 
=- \frac12 \frac{\varepsilon^{\alpha \mu} \varepsilon^{\beta \nu}}{e}
\eta_{ab}e^a_{\mu}e^b_{\nu}\partial_{\alpha}S \partial_{\beta} S \quad .
\end{equation}
The most general dilaton gravity action (\ref{lbegin}) contains the term
$U(X)(\nabla X)^2$. This term can be removed by a dilaton
dependent conformal redefinition of the metric. The matter
action (\ref{lmatter}) is invariant under such a redefinition. However,
the quantum theory is changed: The source terms in eq.\ 
(\ref{source2}) 
below acquire field dependent (conformal) factors, destroying straightforward
quantum integrability. In addition the
path integral measure for the scalar field is
changed. Here, for technical reasons, we restrict ourselves to a
subclass of 2d models with $U(X)=0$. 
Of course, in this way realistic models like spherically symmetric 4d general
relativity \cite{lau}  $U(X) \propto X^{-1}$ are eliminated.

The Lagrangian ${\cal{L}}_s$ containing the source terms for our fields is 
now given by
\begin{equation}
  \plabel{source2}
  {\cal{L}}_s = j^+e_1^- + j^-e_1^+ + j\omega_1 +J^+X^- +J^-X^+ +JX +QS
\end{equation}
Using again an Eddington Finkelstein gauge for the metric defined by
a temporal (Weyl type) gauge for the Cartan variables (\ref{gaugefix})
yields the trivial Faddeev-Popov determinant (\ref{FPdet}).
In this gauge the actions (\ref{lagr}) and (\ref{lmatter}) are
\bea
\plabel{lfirst-gf}
{\cal L}^g_{gf}&=& X^+\pz \emo + X^- \pz \epo +X\pz \omo +X^+ \omo -\epo V(X) \\
{\cal L}^m_{gf}&=& \left( \emo (\pz S)^2 -(\pz S)(\po S)\right) \quad .
\eea
Contrary to the situation in conformal gauge the matter action therefore
still contains a coupling to a zweibein component.
The generating functional of Green functions is defined by
\begin{equation}
\plabel{Zbegin}
W=\int ({\cal D}\sqrt{\epo}S)({\cal D}X)({\cal D}X^+)({\cal D}X^-)
({\cal D}e^+_1)({\cal D}e^-_1)({\cal D}\omega_1)F\exp
\left[\frac{i}{\hbar}\int_x {\cal{L}}_{gf} \right] .
\end{equation}
Note that for the scalar field a nontrivial measure must be introduced in order
to retain invariance under general coordinate transformations \cite{PvN}.
To compute (\ref{Zbegin}) 
we {\it first} integrate here over $\emo$, $X^-$ and $\omo$ to get 
delta functions which are immediately used to integrate out the remaining 
variables $X^+$,
$\epo$ and $X$.
This reduces (\ref{Zbegin}) to
\begin{equation}
W=\int ({\cal D}\sqrt{e^+_1}S) 
e^{\frac{i}{\hbar}\int d^2x (J^-X^++j^-\epo +JX -\epo V(X)-(\pz S)(\po S))} \quad ,
\end{equation} 
where $X^+$, $\epo$ and $X$ thus are expressed as
\begin{equation}
\begin{array}{ccccc}
X^+&=&\pzinv j^+ + \pzinv (\pz S)^2 &=&X_0^+ + \pzinv (\pz S)^2\\
\epo &=&-\pzinv J^+&{}&{} \\
X&=&\pzinv(X^+ +j)&=&X_0 + \pzinvsq (\pz S)^2 \quad .
\plabel{XeX}
\end{array}
\end{equation}
$X_0$ and $X_0^+$ represent $X$ and $X^+$ in the absence of matter fields 
(zero loop order).
The nonlocal expressions for the Green functions $\pz^{-1}$ and $\pz^{-2}$
are regularized as in (\ref{ifcut}). 
$V(X)$ is expanded around $X_0$ 
\begin{eqnarray}
V(X)&=&V_{(0)}+V_{(1)}+\Delta V \\
V_{(0)}&=&V(X_0) \\
V_{(1)}&=&V'(X_0) \pzinvsq (\pz S)^2 \\
\Delta V&=&
\sum_{n=2}^{\infty}\frac{V^{[n]}(X_0)}{n!}\left(\pzinvsq (\pz S)^2 \right)^n
\end{eqnarray} 
The matter field integration to arbitrary orders is contained in the factor
$W_S$ of 
\begin{equation}
\plabel{W-mat}
W=W_S e^{\frac{i}{\hbar}\int d^2x (J^-X_0^++j^-\epo +JX_0 -\epo V_{(0)}}) \quad , 
\end{equation}
i.e.
\begin{equation}
W_S=\int ({\cal D}\sqrt{e^+_1}S) 
e^{\frac{i}{\hbar}\int d^2x  
\left( -\epo \Delta V + (\Emo (\pz S)^2 - (\pz S)(\po S)) -QS \right) }
\end{equation}
We introduced
\begin{equation}
\Emo = \pzinvsq J -\pzinv J^- -\pzinvsq (\epo V'(X_0)) 
\end{equation}
in order to subsume $V_{(1)}$ into the propagator term.
$\Emo$ clearly is not a zweibein component but will formally play a similar 
role.

The integration of the term quadratic in $S$ and thus comprising the full 
propagator in the geometric background
is given by
\begin{equation}
\int ({\cal D}\sqrt{e^+_1}S)
e^{\frac{i}{\hbar} \int d^2x \Emo (\pz S)^2 - (\pz S)(\po S)-QS}=
e^{i\int d^2x S_P(\Emo ,\epo )}e^{\frac{-i}{4\hbar}\int Q\prop^{-1}Q}
\end{equation}
where $\prop^{-1}$ is defined as the inverse of the differential operator
\begin{equation}
\plabel{Gamma}
\prop = \pz \po -\pz E_1^- \pz \quad .
\end{equation}
With a properly regularized $\partial^{-1}_{\mu}$  we assume
an appropriate definition of $\prop$ such that 
$\pz \prop^{-1}= (\po -\Emo \pz )^{-1}$
holds. $S_P$ denotes the Polyakov-Liouville action
\begin{equation}
\plabel{S-Pol}
S_P= \sqrt{-g} R \frac{1}{\square} R 
\end{equation}
where, however,
$R$ and $\square$ have to be expressed in terms of $\Emo$: 
\begin{eqnarray}
{}&{}& e^{\frac{i}{\hbar}
\int (-\epo  \Delta V)}e^{\frac{-i}{4\hbar}\int Q \prop^{-1}Q}
\nonumber \\
{}&{}& =\left( 1 - \frac{i}{\hbar}\int(\epo 
\; \frac{V^{''}(X_0)}{2}\left(\pzinvsq (\pz \frac{i\hbar 
\delta}{\delta Q} )^2 \right)^2 )+...\right)
e^{\frac{-i}{2\hbar}\int Q \prop^{-1}Q} |_{Q=0} \nonumber\\
{}&{}& =1+\int_z i\hbar \epo \frac{V''(X_0)}{8} \gamma (z) +O(\hbar^2)
\end{eqnarray}
where we introduced the abbreviation $\gamma = \gamma (z)$ in the last 
line.  It is now rather straightforward \cite{KLV3} to show that the two loop 
contribution $\gamma$ is {\sl independent} of the fields and in  
the generating functional for connected Green functions 
\begin{eqnarray}
Z&=&\frac{\hbar}{i}\log W  \nonumber \\
\plabel{Z-loop}
{}&=& J^-X_0^++j^- \epo +JX_0 -\epo V_R(X_0)+\hbar S_P(\Emo ,\epo ) +O(\hbar^3)
\\
V_R&=&V-\hbar^2 \gamma V'' \nonumber
\end{eqnarray}
the $\hbar^2$ term expresses a renormalization of the 'potential' 
$V$. Of course, $X_0, X_0^+, e_1^+ $ etc. are expressed in terms 
of the sources.
Let us decompose the potential $V(X)$ in power series of $X$:
\begin{equation}
V(X)=\sum_n \frac {v_n}{n!} X^n
\end{equation}
Any coefficient gets infinite renormalization $\delta v_n
\propto v_{n-2}$. In general, to fix the potential $V$
one needs an infinite number of normalization conditions.
This is not a surprise, however, because even at the classical
level an arbitrary function of $X$ is specified by an infinite
number of independent parameters. There is an important
particular case $V(X)=\alpha \exp (\beta X)$ when the
renormalized potential will be automatically exponential,
and only one parameter $\alpha$ needs to be renormalized.
Note, that this potential gives black hole solutions
\cite{cruz}.

As a final remark to this section we may add that in the 
effective action the Polyakov term does not depend on $E_1^-$ but 
on $e_1^-$, thus does not acquire 2-loop corrections. 

\section{ Hawking radiation for generalized dilaton theories}

This section will return to the more widespread techniques for 
treating radiation of matter from a background with fixed 
singularity structure, preferably from the SSG black hole 
itself.

\subsection{Minimally coupled matter}

The two most frequently considered theories, the string inspired CGHS 
\cite{wit91} and SSG differ drastically in some of their physical 
properties, e.g.\  with respect to the completeness of null geodesics for 
these  
two models \cite{kat97}. These differences directly lead one to investigate
physical properties of a generalized 
model of which the two prominent examples are simply particular cases.

An important feature in semiclassical considerations is the
behavior of Hawking radiation.  In the CGHS model it is just
proportional to the cosmological constant whereas the dependence
in SRG is inverse to its mass, which implies an accelerated
evaporation towards the end of its lifetime.  As we will show a
generalized theory with minimally coupled matter will exhibit
Hawking radiation which is proportional to the black hole mass
in terms of positive or negative powers of the black hole mass,
depending on the parameters of the model \cite{lieblvassil}.  

There are a number of ways of calculating the Hawking radiation
\cite{rev}. One of them consists in comparing  vacua before
and after the formation of a black hole. In the case of generalized
dilaton gravity this way is technically rather involved. We
prefer a simpler approach based on an analysis of static black
hole solutions.

Consider a generalized Schwarzschild black hole given by
\begin{equation}
ds^2=-L(U)d\tau^2+L(U)^{-1}dU^2 , \label{gSch}
\end{equation}
where $L(U)$ has a fixed behavior at the asymptotic region ${\cal I}^+$:
\begin{equation}
L(U) \to L_0(U) \label{Jas}
\end{equation}
with $L_0(U)$ corresponding to the ground state solution.
At the horizon we have $L(U_h)=0$. We can calculate the geometric Hawking
temperature as the normal derivative of the norm of the Killing vector
$\partial /\partial \tau$ at the (nondegenerate) horizon 
\begin{equation}
T_H=|\frac 12 L'(U_h) |. \label{THgen}
\end{equation}

We introduce the coordinate $z$ by 
\begin{equation}
dU =dz L(U). \label{Jdefz}
\end{equation}
Then the metric takes the conformaly trivial form with
\begin{equation}
ds^2 = e^{2\rho} (-d\tau^2+dz^2) \qquad \rho =\frac 12 \ln L. \label{Jdefrho}
\end{equation}
In conformal coordinates the stress energy tensor looks
like \cite{rev}
\begin{equation}
T_{--}=-\frac 1{12\pi} ((\partial_-\rho )^2 -
\partial_-\partial_-\rho )+t_- =T_{--}[\rho (L)]+t_-.
\label{JTmm}
\end{equation}
One can choose coordinates such that in the asymptotic
region
\begin{equation}
T_{--}[\rho (L_0)]=0 . \label{TJ0}
\end{equation}
This choice ensures that
there is no radiation in the ground state. It means
that we measure Hawking radiation of a black hole without any contribution
from background Unruh radiation.

The constant $t_-$ is defined by the condition at the horizon
\begin{equation}
T_{--}|_{\rm hor}=0  \label{Thor}
\end{equation}
in the spirit of  \cite{CrFu}. 
The corresponding vacuum state is called the Unruh vacuum.
In this state there is no energy flux at the black hole horizon.
Of course, there cannot be such a thing as an observer at
the horizon. However, as we shall demonstrate bellow, predictions of the 
theory with regard to measurements made at infinity are independent 
of the choice of coordinates at the horizon. In the case of four
dimensional black holes this is well known.
Taking into account equations
(\ref{JTmm}), (\ref{TJ0}) and (\ref{Thor}) one obtains a relation
between $T_{--}[\rho ]$ at the horizon and the asymptotic value
of $T_{--}$:
\begin{equation}
T_{--}|_{\rm asymp}=-T_{--}[\rho ]|_{\rm hor} \label{ashor}
\end{equation}
The following simple identities are useful:
\begin{equation}
\partial_-=\frac 12 \partial_z ,\quad
\partial_z \rho = \frac 12 L' , \quad
\partial_z^2 \rho = \frac 12 L'' L ,
\label{useful}
\end{equation}
where prime denotes differentiation of $L$ with respect to
$U$. By substituting (\ref{useful}) into (\ref{ashor}) we obtain
the Hawking flux
\begin{equation}
T_{--}|_{\rm asymp}=\frac 1{48\pi} (\frac 12 L'(U))^2|_{hor}.
\label{Jflux}
\end{equation}

It is easy to demonstrate that our result is independent of a particular
choice of conformal coordinates provided the behavior at the asymptotic
region is fixed.  For the models with the background given by the 
solutions of (39) we obtain the Hawking flux for the subclass $b 
= a - 1 \neq 0$ (asymptotic Minkowski-spacetime)

\begin{equation}
T_{--}|_{\rm asymp}=\frac{a^2}{384\pi} 
C^{\frac{2(a-1)}{a}}\left( \frac{2B}{a} \right)^{\frac{2-a}{a}}\; ,
\end{equation}
to be compared with CGHS $(a = 1)$

\begin{equation}
T_{--}|_{\rm asymp}=\frac{B}{192\pi}  = \frac{\lambda^2}{48\pi}\; .
\end{equation}

\subsection{Nonminimally coupled scalars}

Somewhat surprisingly until very recent times 
\cite{bousso,noj97a,ElNO,mik97,noj97b,KLV2} no 
computation of the Hawking radiation for that case seems to 
exist. The purpose of this section is to give a comprehensive 
and direct answer to that question including all physically 
interesting models which generalize spherically symmetric gravity 
(SSG). This also allows us to improve and correct the results of 
\cite{bousso,noj97a,ElNO,mik97,noj97b} and to 
show the arbitrariness involved when SSG is generalized.

In the SSG case the (ultralocal) measure for the matter 
integration is well defined, because

\begin{equation}
\plabel{ca1}
\int d\Omega \sqrt{-^4g} =  e^{-2\phi}\sqrt{-g} \quad ,
\end{equation}

For the generalized class of models (\ref{lbegin}), however, this definition 
is not unique, as well as the one for an eventual nonminimal 
factor for the possible coupling to matter in (\ref{lmatter}). Therefore, in 
that case we have to allow the general replacements $\Phi \to 
\varphi (\Phi)$ in the SSG-factor $e^{-2\phi}$ for (\ref{lmatter})
 and $\Phi \to \psi (\Phi) $ in 
(\ref{ca1}), where 
$\varphi$ and $\psi$ may be general (scalar) functions of the 
dilaton field. With these replacements and in terms of the field 
$\tilde f = f\; e^{-\psi}$ which satisfies the standard 
normalization condition, (\ref{lmatter}) can be rewritten as

\begin{equation}
\plabel{tact}
S = -\frac 12 \int \sqrt{-g}d^2x 
\tilde f A \tilde f
\end{equation}

where
\begin{equation}
A = -e^{-2\varphi +2\psi}g^{\mu\nu} (\nabla_\mu \nabla_\nu
+2(\psi_{,\mu}-\varphi_{,\mu})\nabla_\nu +\psi_{,\mu\nu}
-2\varphi_{,\mu} \psi_{,\nu} ), \plabel{A}
\end{equation}

The path integral for $\tilde f$ leads to the effective action 

\begin{equation}
W = \frac 12  \mbox{Tr} \ln A \quad .
\end{equation}

After continuation to the Euclidean domain $A$ becomes an 
elliptic second order differential operator.
The corresponding one loop effective action $W$ can be expressed 
in terms of the zeta function of the operator $A$:\footnote{For an 
extensive discussion of that technique consult \cite{esposito}.}

\begin{equation}
W=-\frac 12 \zeta'_A(0), \qquad \zeta_A(s)={\rm Tr}(A^{-s})
\plabel{zeta}
\end{equation}
Prime denotes differentiation with respect to $s$. From $W$ 
regularized in this way an infinitesimal conformal transformation 
$\delta g_{\mu\nu} = \delta k g_{\mu\nu}$ produces the 
trace of the (effective) energy momentum tensor

\begin{equation}
\delta W=\frac 12 \int d^x \sqrt g\delta g^{\mu\nu}T_{\mu\nu}
=-\frac 12 \int d^x \sqrt g \delta k(x)T_\mu^\mu (x)
\plabel{T}
\end{equation}

Due to the transformation property $\delta A = -\delta k A$ 
of (6) (valid in $d = 2$ only) with the definition of a 
generalized $\zeta$-function

\begin{equation}
\zeta (s|\delta k,A)={\rm Tr}(\delta kA^{-s})
\plabel{varW}
\end{equation}

the variation in (\ref{T}) can be identified with 

\begin{equation}
\plabel{TX}
\delta W=-\frac 12 \zeta (0|\delta k,A)
\end{equation}

Combining (\ref{TX}) and (\ref{T})  we get

\begin{equation}
\zeta (0|\delta k,A)=\int d^x \sqrt g \delta 
k(x)T_\mu^\mu (x)\; .
\plabel{T2}
\end{equation}

By using the Mellin transformation one can show that 
$\zeta (0|\delta k,A)=a_1(\delta k,A)$ \cite{gil75}, where $a_1$ is
defined as a coefficient in small $t$ 
asymptotic expansion of the heat kernel:
\begin{equation}
{\rm Tr}(F\exp (-At)) =\sum_n a_n (F,A)t^{n-1}
\plabel{hk}
\end{equation}
To evaluate $a_1$ we use the standard method \cite{gil75}.
To this end we represent $A$ as
\begin{equation}
A=-(\hat g^{\mu\nu}D_\mu D_\nu +E) ,\qquad 
E=\hat g^{\mu\nu}(-\varphi_{,mu}\varphi_{,\nu}
+\varphi_{,\mu\nu}) \plabel{newA}
\end{equation}
where $\hat g^{\mu\nu}=e^{-2\varphi +2\psi}g^{\mu\nu}$,
$D_\mu =\nabla_\mu +\omega_\mu$, $\omega_\mu =\psi_{,\mu}-
\varphi_{,\mu}$. For $a_1$ follows \cite{gil75}:
\begin{equation}
a_1 (\delta k, A)=\frac 1{24\pi} \int d^2x\sqrt{-\hat g} \delta k  
(\hat R+6E)\; .
\plabel{a1}
\end{equation}
Returning to the initial metric and comparing with (\ref{T}) 
we obtain the most general form of the conformal anomaly
\begin{equation}
T_\mu^\mu =\frac 1{24\pi} (R-6(\nabla \varphi )^2 +
4\square \varphi +2\square \psi )
\plabel{T3}
\end{equation}

It is not difficult to adapt the methods of section 4.1 to obtain 
the Hawking flux also here. In the CGHS the result is

\begin{equation}
T^{CGHS}_{--}\vert_{asymp}=\frac{\lambda^2}{48\pi}(1+
\frac 32 \alpha^2-2\alpha -\beta)
\plabel{CGHS}
\end{equation}

Even for minimal coupling ($\alpha = 0$) this expression is 
inherently ambiguous due to the constant $\beta$ which had its 
roots in the ambiguous definition of an ultralocal measure. 
Increasing $\alpha$ (nonminimal coupling) above $\alpha = 4/3$ 
tends to increase $T_{--}$. Of course, by adjusting $\beta$ the 
flux may become zero or even negative as well ('cold dilaton 
black hole'). Like the (geometric) Hawking temperature (\ref{CGHS}) in 
this case does not depend on the mass of the black hole. 

The final result for $a \neq 1$ reads

\begin{equation}
T^{(a)}_{--}\vert_{asymp}=
\frac{1}{48 \pi}T_H^2
\left ( 1-\frac{3\alpha^2}{2(2-a)} -
\frac{1}{2-a} (2\alpha +\beta ) \right ) 
\plabel{Ta}
\end{equation}

which has been expressed in terms of the geometric 
Hawking temperature for the general models \cite{admbondi} 

\begin{equation}
T^2_H = \frac{a^2}{8} C^{\frac{2(a-1)}{a}} 
\left(\frac{2B}{a}\right)^{\frac{2-a}{a}}
\end{equation}

For SSG all parameters are unambiguously given ($a = \frac 12, \alpha = 
\beta = 1$). Then the bracket in (25) yields a factor $-2$, i.e.\ a 
negative flux! For a special case the same qualitative result has been obtained
already in \cite{wipf}. 

The crucial difference of our method is the use of a {\it local} 
scale transformation inside the zeta function. Due to the 
presence of an arbitrary {\it function} $\delta\, k$ 
all terms there are fixed unambiguously. 

Our result for the 'anomaly', (\ref{T3}), is the most general one 
obtainable in 1+1 dimensional theories. The result for the 
Hawking flux in SSG, on the other hand, taken literally would 
mean that an influx of matter is necessary to maintain in a kind 
of thermodynamical equilibrium the Hawking temperature of a black 
hole --- in complete contradiction to established black hole 
wisdom. However, to put this result on a sound basis the 
treatment of Hawking radiation in the asymptotic region in that 
case certainly requires to go beyond the usual approach adopted 
also in our present paper. After all, non-minimally coupled 
scalar fields are strongly coupled in the asymptotic region. 
Therefore a result like (\ref{Ta}) for SSG cannot be the final answer. 
In fact, probably new methods for extracting the flux towards 
infinity in such a case have to be invented. 

\section{Acknowledgement}

This work has been supported by Fonds zur F\"orderung der 
wissenschaftlichen Forschung (FWF) Project No.\ P 10221-PHY. One 
of the authors (D.V.) thanks GRACENAS and the Russian Foundation 
for Fundamental Research, grant 97-01-01186, for financial 
support. Another author (M.O.\ K.) is grateful to the Erwin 
Schr\"odinger International Institute, The International Science 
Foundation (grant NFR 000) The Russian Fund for Fundamental 
Investigations (grant RFFI-96-010-0312) and the Austrian Academy 
of Sciences.

\vfil


\begin{thebibliography}{99}
\bibitem{very}A very comprehensive global analysis of generally 
                covariant 2d models, without concentrating especially on 
                dilaton theories, has only appeared recently \cite{alt}, whereas 
                models with torsion had been discussed in detail already 
                before \cite{kat93}.
\end{thebibliography}
\end{document}